\documentclass[a4paper,twocolumn,11pt,accepted=2025-12-08]{quantumarticle}
\pdfoutput=1
\usepackage[utf8]{inputenc}
\usepackage[english]{babel}
\usepackage[T1]{fontenc}
\usepackage{amsmath}
\usepackage{hyperref}

\usepackage{tikz}
\usepackage{lipsum}

\usepackage{multicol}
\usepackage[justification=centerlast]{caption}
\usepackage[labelformat=simple]{subcaption}

    \usepackage[justification=centerlast]{caption}

\usepackage{xcolor}

\begin{document}

\title{Quantum Communication Networks Enhanced by Distributed Quantum Memories}

\author{Xiangyi~Meng}%
\affiliation{Department of Physics, Applied Physics, and Astronomy, Rensselaer Polytechnic Institute, Troy, New York 12180, USA}%
\affiliation{Network Science and Technology Center, Rensselaer Polytechnic Institute, Troy, New York 12180, USA}%
\affiliation{Department of Physics and Astronomy, Northwestern University, Evanston, Illinois 60208, USA}%
\orcid{0000-0001-5184-7648}

\author{Nicolò~Lo~Piparo}%
\affiliation{Okinawa Institute of Science and Technology Graduate University,
1919-1 Tancha, Onna-son, Okinawa 904-0495, Japan}%

\author{Kae~Nemoto}%
\affiliation{Okinawa Institute of Science and Technology Graduate University,
1919-1 Tancha, Onna-son, Okinawa 904-0495, Japan}%

\author{István~A.~Kovács}%
\email{istvan.kovacs@northwestern.edu}
\affiliation{Department of Physics and Astronomy, Northwestern University, Evanston, Illinois 60208, USA}%
\affiliation{Northwestern Institute on Complex Systems, Northwestern University, Evanston, Illinois 60208, USA}
\affiliation{Department of Engineering Sciences and Applied Mathematics, Northwestern University, Evanston, Illinois 60208, USA}

\maketitle

\onecolumngrid
\begin{abstract}
Building large-scale quantum communication networks has its unique challenges. Here, we demonstrate that a network-wide synergistic usage of quantum memories distributed in a quantum communication network offers a fundamental advantage. We first map the problem of quantum communication with local usage of memories into a classical continuum percolation model. Then, we show that this mapping can be improved through a cooperation of quantum distillation and relay protocols via remote access to distributed memories. This improved mapping, which we term $\alpha$-percolation, can be formulated in terms of graph-merging rules, analogous to the decimation rules of the renormalization group treatment of disordered quantum magnets. These rules can be performed in any order, yielding the same optimal result that is characterized by the emergence of a ``positive feedback'' mechanism and the formation of spatially disconnected ``hopping'' communication components---both marking significant improvements beyond the traditional point-to-point consideration of quantum communication in networked structures.\newline
\end{abstract}

\twocolumngrid


\begin{center}
\begin{table*}[t!]
	\centering
        \vspace{0cm}
	\caption{\label{table_representation}{Glossary of problem representations. We map generic quantum communication protocols (first row) to two graph-merging rules (second row), which give rise to an $\alpha$-percolation process with enhanced features (third row).}\hfill\hfill}

        \fontencoding{T1}
        \fontsize{8}{10}\selectfont
		{\begin{tabular}{ccc}
		\hline\hline
        Variables: $r_a,d_{bc}$ & $(r')^{1/\alpha}=r_a^{1/\alpha}+r_b^{1/\alpha}$ [Eq.~\eqref{eq_contraction_r}] & $d'_{bc}=\min\{d_{bc}, d_{ab}+d_{ac}\}$ [Eq.~\eqref{eq_reduction}] \\
        \hline
		Quantum communication protocol (Fig.~\ref{fig_3d})& 
        remote distillation & 
        quantum relay\\
		Graph-merging rules [Figs.~\ref{fig_rule_a}~and~\ref{fig_rule_b}] & contraction &  
            reduction\\
		$\alpha$-percolation [Fig.~\ref{fig_demo}] & positive feedback 
        & hopping\\
		\hline\hline
	\end{tabular}}
        \vspace{0cm}
\end{table*}
\end{center}


\section{Introduction}

Rapid advancements in {quantum information} technologies call for a {proactive} understanding of the potential benefits of using \emph{quantum memory} elements, a crucial resource constraint~\cite{q-repeater_rpl09,q-mem_agr15,q-repeater_jr22,q-mem_mgghjckclr23,q-mem_gwfsnkssvvf24}. 
Quantum memories, akin to their classical counterparts, 
play an essential role in enabling
complex information processing. For example, in quantum machine learning, certain algorithms leveraging quantum memories require an exponentially smaller number of runs to achieve a given accuracy~\cite{qml_bwprwl17,qml_hkp21}. Similarly, it has been shown that simulating stochastic processes with arbitrary precision is possible with  significantly fewer memories in a quantum setting~\cite{q-mem_pghwp17,q-mem_eg18}.

{In quantum communication~\cite{q-commun_gt07,q-commun_yblzpp10,q-internet_weh18},}
{a primary  usage of quantum memories is for {distillation protocols}~\cite{q-distill_hh01,entangle-purif_db07,q-distill_ascfs16}, aiming to generate a high-fidelity entangled pair for long-range communication by ``distilling'' from many low-fidelity pairs.
A distillation protocol typically consists of many one-shot purification steps. Each purification step, designed to transform two less entangled pairs 
into a single strongly entangled pair, 
only succeeds with success probability $p<1$~\cite{q-distill_hh01,entangle-purif_db07,q-distill_ascfs16}.
This probabilistic nature
poses considerable complexity in scaling up to $n$ pairs: without quantum memories, all $O(n)$ {purification steps} must succeed simultaneously. As a result, the average waiting time to witness a successful outcome extends exponentially to $O(p^{-n})$, assuming that $p$ is the same for all {steps}. On the other hand, with sufficient $O(n)$ memories, all purification steps can be executed in a nested way, by distilling $n/2$ pairs from $n$ pairs in parallel each time.}
This way, the whole process takes only $O(\log_2 n)$ {rounds}, significantly reducing the time complexity to $O(p^{-\log_2 n})=O(n^{-\log_2 p})$, a polynomial in $n$~\cite{q-distill_hh01,entangle-purif_db07,q-distill_ascfs16}.

This example illustrates that
entanglement can be 
more efficiently generated point-to-point between parties equipped with memories, yielding longer communication ranges~\cite{q-repeater_rpl09,q-mem_agr15,q-repeater_jr22,q-mem_mgghjckclr23,q-mem_gwfsnkssvvf24}.
{However, as we will show, this enhancement extends beyond point-to-point scenarios.
Consider a spatial network, where memories are \emph{distributed} across multiple parties (nodes), not just two, resembling modern distributed computing architectures~\cite{could-comput_bgmfs18}.
It is conceivable  that 
each node can utilize not only their own memories to generate entanglement but also the remote memories, accessing them through existing quantum channels.} This raises a critical question: \emph{can distributed quantum memories offer major enhancements (beyond point-to-point) in quantum communication networks?} Surprisingly, despite the rich literature on quantum networks~\cite{q-netw_wjzlyflzwdsogz22,conpt_mhtdlgh23,q-netw_npb23}, there lacks a general network model for distributed quantum memories, which should be characterized as node weights in the network. This sets the model apart from other well-developed link-weighted quantum network models, such as photonic communication~\cite{q-netw_bccc20,q-netw_bccc21}, channel capacity bounds~\cite{q-channel_pblocsb18,q-netw-route_p19}, or entanglement percolation theories~\cite{conpt_mgh21,conpt_lcw22,conpt_mmhksg22,det_mcghr23}.

In this work, we bridge this gap by first 
mapping the problem of quantum communication with distributed memories to a classical {continuum percolation} model~\cite{contin-percolation}, {where the number of memories at each node $a$ directly maps to the node's ``range,'' denoted by $r_a$. 
In our model, to demonstrate a fundamental mechanism in its purest form, we consider ideal quantum memories with infinite coherence time, error-free operations, and perfect transduction efficiency. Consequently, the quantum advantage provided by each node $a$ is fully characterized by the quantity $r_a$. This idealization allows us to focus on the network aspects of the model rather than underlying engineering challenges~\cite{q-repeater_rpl09}. 
The continuous percolation model also defines a distance $d_{ab}$ between every two nodes $a$ and $b$, quantifying the difficulty of distilling entanglement between them. Note that $d_{ab}$ can be an effective distance that does not necessarily correspond to physical separation. With these definitions, we find the intuitive result that quantum communication is only possible between $a$ and $b$ when they fall into each other's range and ``connect,'' namely $d_{ab}<\min\{r_a, r_b\}$.
This threshold represents the condition that there are sufficient memories, in both $a$ and $b$, that can be used to distill entanglement across the distance $d_{ab}$ efficiently (i.e.,~with acceptably high fidelity and sub-exponential time complexity).}

{Within this setting, our main result is the finding that nodes can achieve even greater ranges through a protocol we call \emph{remote distillation}, which outperforms point-to-point distillation by 
also utilizing---through quantum gate teleportation~\cite{q-gate-teleport_jtsl07}---the memories of other nodes that are elsewhere connected to the two communicating endpoints.
In addition to increasing the nodes' ranges, 
the distances between nodes can also be reduced, facilitated by the \emph{quantum relay} protocol~\cite{q-relay_jpf02,q-relay_rmtzcg04}.
Together, the remote distillation and quantum relay protocols map to
two graph-merging rules, 
which can iteratively increase the range $r_a$ and decrease the distance $d_{ab}$, respectively (Table~\ref{table_representation}).
We derive that the order in which these rules are applied does not affect the final, optimal communication configuration.
This resulting (history invariant) configuration maps to an enhanced model we term \emph{$\alpha$-percolation}, characterized by a single efficiency parameter, $\alpha$. The value of $\alpha$ depends on the specific distillation protocol employed and quantifies its efficiency.
We show that this framework substantially enhances network connectivity, a result we demonstrate using both model and real-world fiber network topologies.

{The rest of the paper is organized as follows. We begin in Sec.~2 by establishing the continuum percolation model in detail. In Sec.~3, we enhance this model by incorporating remote distillation and quantum relay protocols. These protocols give rise to the resulting $\alpha$-percolation, which we formalize in Sec.~4. We then present the model's practical utility in Sec.~5. We conclude in Sec.~6 with a discussion of the model's limitations and avenues for future research.}

\section{Continuum Percolation Mapping}

We start by considering a collection of $N$ nodes \emph{without} memories, each spatially embedded such that a distance $d_{ij}$ is defined between every two nodes $i$ and $j$, measuring for instance the length of optical fibers connecting the nodes. 
{Each node uses flying qubits (e.g.,~photons) to establish bipartite entanglement with other nodes. These flying qubits are considered to be cheap and abundant in quantity (assuming they allow heralding~\cite{q-repeat_mlkllj16,q-netw_hkmsvtmh18}), but compromised in quality when they transmit through noisy, distance-dependent quantum channels.} 
We model these channels by the {depolarizing} channel, a universal, worst-case noise model~\cite{werner_y02,q-depolarize_k03}.
The outcome of the depolarizing channel is a bipartite mixed isotropic state (Werner state)~\cite{werner_w89} that links nodes $i$ and $j$, described by the density matrix,
\begin{equation}
\label{eq_werner}
   \rho (d_{ij})= p(d_{ij})\left|\Psi^{-}\right\rangle\left\langle\Psi^{-}\right|+\left(1-p(d_{ij})\right) \mathbf{I}/4.
\end{equation}
Here, $\mathbf{I}$ is the identity matrix, representing the state of maximum noise, and $\left|\Psi^{-}\right\rangle\left\langle\Psi^{-}\right|$ is a maximally entangled state between two qubits, specifically one of the four Bell states.
The distance dependence of the coefficient $p(d_{ij})$ is modeled by an exponential decay, $p(d_{ij})=e^{-d_{ij}/d_0}$,
where $d_0$ signifies the characteristic decoherence distance, 
dictated by current technological constraints and limited to hundreds of kilometers~\cite{q-netw-200km_dtyshhktnas09,q-netw-fiber_g16}.

For practical quantum communication, the fidelity between the noisy state $\rho (d_{ij})$ and the ideal state $\left|\Psi^{-}\right\rangle\left\langle\Psi^{-}\right|$, given by
$F(d_{ij})=\left(3 p (d_{ij})+1\right)/4$,
 must be sufficiently close to unity. We thus seek to reach a fidelity threshold
$F_\text{th}\equiv 1-\epsilon$ where $\epsilon$ represents a small error typically required to be smaller than $1\%$~\cite{q-comput_ebkemzlgwmlsgvl23}.
This allows us to introduce a uniform, constant  range for every node, $r_0=-d_0\ln \left(1-4\epsilon/3\right)$, such that two nodes can directly establish usable entanglement only if $d_{ij}<r_0$.
For $\epsilon\ll 1$ this formula simplifies to $r_0\simeq 4\epsilon d_0 /3$.
When $d_{ij}>r_0$, the level of entanglement is insufficient to meet the fidelity threshold ($F(d_{ij})<F_\text{th}$).
This consideration thus maps to {continuum percolation}: around each node $i$, a disk
with a fixed, uniform radius $r_0$ is placed; only when two disks can reach each other's center\footnote{Here, our ``center-reaching'' convention ($d<r_0$) differs from the traditional ``periphery-reaching'' convention ($d<2 r_0$), which can be achieved by replacing our $r_0$ by $2 r_0$. However, only the center-reaching convention upholds the geometric interpretation of Eq.~\eqref{eq_criterion_mega} once moving to $\alpha$-percolation.}, 
they forge a link, which results in a connected component.

\begin{figure}[t!]
	\centering
	\hspace{-10mm}
    \begin{minipage}[b]{130pt}	
		{\includegraphics[width=130pt]{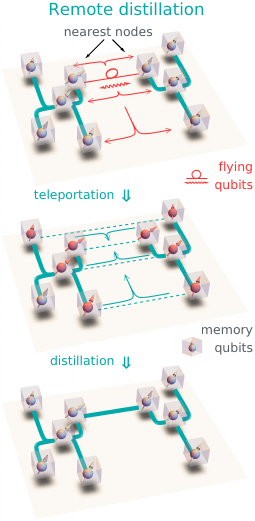}\vspace{-4mm}\subcaption{\label{fig_3d_distill}}}
    \end{minipage}
    \begin{minipage}[b]{100pt}	
		{\includegraphics[width=130pt]{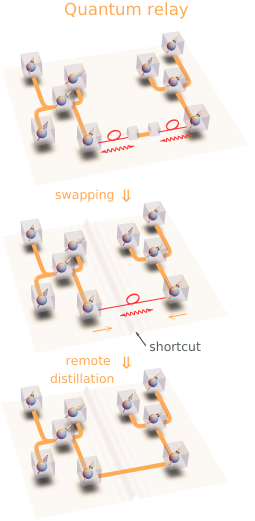}\vspace{-4mm}\subcaption{\label{fig_3d_relay}}}
	\end{minipage}
	\caption{\label{fig_3d}Quantum communication enhanced by distributed memories. \subref{fig_3d_distill}~Quantum memories distributed across two connected components (the X- and Y-shaped) effectively act as ``cloud storage'' for remotely storing flying qubits generated between the nearest nodes of the two components. This allows distillation between the two components, potentially forging a new entanglement link ($F>F_\text{th}$). \subref{fig_3d_relay}~A connected component can also act as a relay to swap flying qubits between its neighbors. This ``shortcut'' further promotes remote distillation to forge new links.
 \hfill\hfill}
\end{figure}

\section{Enhancement from Quantum Memories}

Now we consider the case where each node is equipped with $m$ long-life, repeatedly usable {quantum memories} (e.g.,~NMR, defect, or superconducting qubits). 
This already  brings an improvement {at the point-to-point level (Appendix~\ref{appendix_a})}: between every two nodes, a pair of stronger entanglement can now be distilled. 
{Specifically,} when merging two pairs with a fidelity of $F$, utilizing the BBPSSW distillation protocol~\cite{entangle-purif_bbpssw96}, a new pair is created with 
a fidelity of $F'$ that follows
$1-F'\simeq\left(2/3\right)\left(1-F\right)$ in the limit $F\to 1$.
Consequently, with a maximum of $m$ pairs that can be stored, the final fidelity $F'$ scales as $1-F'=\epsilon\simeq\left(2/3\right)^{\log_2 m}\left(1-F\right)$. Therefore, each initial pair before distillation only needs to achieve a fidelity of $1-\left(3/2\right)^{\log_2 m} \epsilon$. This increases each node's range to
\begin{equation}
\label{eq_r_0}
r_0=-d_0\ln\left[1-\frac{4}{3} \epsilon m^{\log_2\left(3/2\right)}\right]\simeq \frac{4}{3}\epsilon m^{\alpha^* }d_0,
\end{equation}
where $\alpha^*\equiv \log_2\left(3/2\right) \approx 0.585$.
As we show next, enhancements are not confined to {the point-to-point level}, but also prevail at the \emph{component} level. As nodes form larger components, there are (at least) two additional improvements:

\subsection{Remote distillation}
{Essentially, remote distillation utilizes teleportation of qubits and gates (which comes with a cost) to perform distillation remotely. This approach leverages the benefits of having quantum memories distributed across the network.}
Let us consider two components $a$ and $b$, with respective {sizes (numbers of nodes)} $s_a$ and $s_b$.
The {total} number of distributed quantum memories within these components are proportional to $m s_a$ and $m s_b$. 
{Assume that flying qubits are established through the {nearest} nodes
between $a$ and $b$~(Appendix~\ref{appendix_b}). Here, ``nearest'' guarantees that the flying qubits have the highest quality to be achieved between $a$ and $b$.
{The links within each component have $F \simeq 1$ and thus can further teleport and store the flying qubits from the nearest node to the other nodes within the same component} [Fig.~\ref{fig_3d_distill}]. 
In other words, the other nodes' memories now effectively function as extra ``cloud storage'' for the flying qubits.}
These stored qubits are further distilled with the help of remote gate teleportation~\cite{q-remote-control_hvcp01}, resulting in a single entangled pair between $a$ and $b$, with an even higher fidelity that scales as $1-F'=\epsilon\simeq\left(2/3\right)^{\log_2 n}\left(1-F\right)$. Here, the total number of stored pairs $n$, {which are to be distilled,} depends on the \emph{smaller} number of available memories in the two components, given by $n=\min\{m s_a, m s_b \}$.
Essentially, this remote distillation allows us to rewrite the connection criterion, $d_{ij}<r_0$, to
\begin{equation}
\label{eq_criterion}
    \min_{i\in a, j\in b}{d_{ij}}<\min\{r(s_a), r(s_b)\},
\end{equation}
where $r(s)$ denotes the improved range of a node in a component of size $s$, calculated as $r(s)\simeq 4\epsilon m^{\alpha } s^{\alpha }d_0/3$, meaning that each node within a component of size $s>1$ has a range exceeding $r_0$ in Eq.~\eqref{eq_r_0}. Here, instead of a fixed $\alpha^*$, we assume a generic value $\alpha>0$.

{The newly forged link between components $a$ and $b$, in turn, allows the sharing of the total memories of \emph{both} $a$ and $b$. Thus, given that the two components are now merged into a larger component of size $s_a+s_b$, every node within this enlarged component will gain a uniform, increased range $r'=r(s_a+s_b)$,} given by 
\begin{equation}
\label{eq_contraction_r}
    (r')^{1/\alpha}=r_a^{1/\alpha}+r_b^{1/\alpha},
\end{equation}
where $r_a\equiv r(s_a)$ and $r_b\equiv r(s_b)$ [Fig.~\ref{fig_rule_a}].

{Remote distillation requires existing links within the components (of $\sim n$ total memories) to facilitate teleportation of flying qubits and remote gates, which costs at most $\sim n^2$ links.
Taking into account the replenishment of these links, our analysis shows that the time complexity of the protocol scales as $f(n)=O(n^{\log_2 n})$ (Appendix~\ref{appendix_c}). While this is greater than the polynomial complexity $O(n^{-\log_2 p})$ of point-to-point distillation (see Introduction), it remains subexponential in $n$, thus underscoring the enhancement at the component level.}

Equivalently, we can view each component as a single effective node $a$ of range $r_a$ and define the distance between two effective nodes as $d_{ab}=\min_{i\in a, j\in b}{d_{ij}}$. This effective-node picture simplifies the criterion Eq.~\eqref{eq_criterion} to
\begin{equation}
\label{eq_criterion_mega}
    d_{ab}<\min\{r_a, r_b\}.
\end{equation}

\begin{figure*}[t!]
	\centering
    \begin{minipage}[b]{90pt}	
		{\includegraphics[width=180pt]{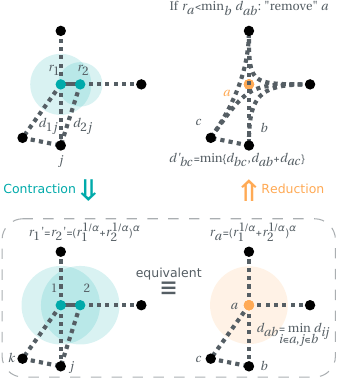}\vspace{-2mm}\subcaption{\label{fig_rule_a}}}
    \end{minipage}
    \begin{minipage}[b]{90pt}	
		{\subcaption{\label{fig_rule_b}}}
    \end{minipage}
    \begin{minipage}[b]{90pt}	
		{\includegraphics[width=90pt]{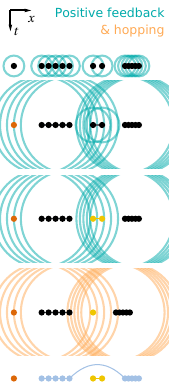}\vspace{-2mm}\subcaption{\label{fig_demo}}}
	\end{minipage}
	\caption{\label{fig_percolation}$\alpha$-percolation with graph rules. \subref{fig_rule_a}~Contraction rule: we redefine a uniform range for every node within the same component [Eq.~\eqref{eq_contraction_r}], which is equivalent to contracting $1,2,\cdots$ into a single ``effective node'' $a$ of range $r_a$. 
    \subref{fig_rule_b}~Reduction rule: we 
    remove an effective node $a$ from percolation but create a shortcut between each pair of its neighbors ($b,c,\cdots$) [Eq.~\eqref{eq_reduction}]. This only happens once $a$ {satisfies} $r_a<\min_b d_{ab}$.
    \subref{fig_demo}~During the $\alpha$-percolation process (from top to bottom), nodes within mutual reach are joined as a larger component, which in turn increases each node's new range {(denoted by the circles' radii)}, representing a ``positive feedback'' mechanism. When a component cannot reach the others (highlighted in distinct colors, e.g.,~the yellow nodes), it can still provide shortcuts that may facilitate ``hopping'' links {(e.g.,~the blue link)} between other components.
 \hfill\hfill}
\end{figure*}

\subsection{Quantum relay}

A {component} can also function as a quantum relay~\cite{q-relay_jpf02,q-relay_rmtzcg04} and swap
{flying} qubits between its nodes [Fig.~\ref{fig_3d_relay}] by entanglement swapping~\cite{entangle-swap_zzhe93}.
Thus, given a component $a$, each pair of its neighboring components (denoted by $b,c$) can form an entangled pair, as if $b$ and $c$ formed a  ``shortcut'' through $a$ [Fig.~\ref{fig_rule_b}].
This relay function does not require, or benefit from, quantum memories~(Appendix~\ref{appendix_d}), but it may enhance further remote distillation by decreasing the distance between every pair of $a$'s neighbors to
\begin{equation}
\label{eq_reduction}
    d'_{bc}=\min\{d_{bc}, d_{ab}+d_{ac}\}.
\end{equation} 
Equation~\eqref{eq_reduction} is derived from the fact that given two entangled pairs, one between $a,b$ and the other between $a,c$, entanglement swapping~\cite{entangle-swap_sdsbbz05} at the effective node $a$ (more precisely, at all nodes $i\in a$ which are along the path connecting $b$ and $c$) gives 
$p'_{bc}=p(d_{ab})p(d_{ac})$ by Eq.~\eqref{eq_werner}, thus yielding the term $d_{ab}+d_{ac}$ in Eq.~\eqref{eq_reduction}.
Moreover, given that flying qubits are abundant, if there is already a channel between $b$ and $c$ (characterized by $d_{bc}$), only the {shortest} link connecting the two components contributes to further distillation processes. This warrants the minimum rule in Eq.~\eqref{eq_reduction}~(Appendix~\ref{appendix_d}).

\begin{figure}[t!]
	\centering
    \begin{minipage}[b]{220pt}	
		{\includegraphics[width=220pt]{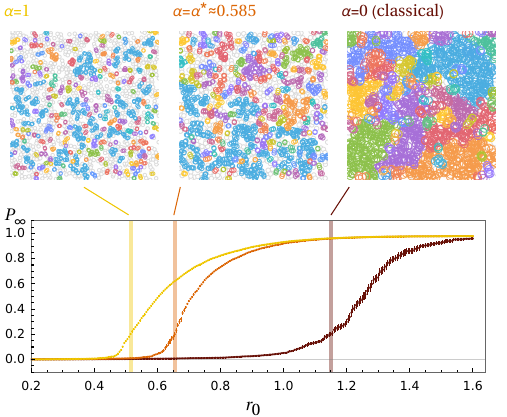}\subcaption{\label{fig_threshold}}}
    \end{minipage}

    \begin{minipage}[b]{220pt}	
		{\includegraphics[width=220pt]{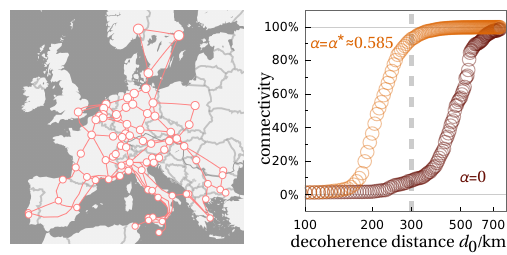}\subcaption{\label{fig_real}}}
    \end{minipage}
	\caption{$\alpha$-percolation on two-dimensional networks.
    \subref{fig_threshold}~Uniformly random points. As the exponent $\alpha$ increases, not only does the threshold $r_0^\text{th}$ decrease (brown to yellow), but also the giant component ({of relative size} $P_{\infty}$) at criticality exhibits a more nonlocal and sparser distribution across space. Here, each node is illustrated using its initial range $r_0$, rather than the actual range $r(s)$. Components with size $s>1$ are highlighted in distinct colors.
    \subref{fig_real}~Pan-European fiber network, consisting of $692$ nodes (including both stations and repeaters) and $733$ links (fiber optic cables). 
    The fidelity bound is set at {$F_\text{th}\equiv1-\epsilon=99\%$}. Each node is assigned $m \approx 102$ quantum memories such that $m^{\alpha^*}\approx m^{0.585}=15$.
    \hfill\hfill}
\end{figure}

\section{Graph Rules}

{From the remote distillation and quantum relay protocols we derive
two rules: the \emph{contraction} [Eq.~\eqref{eq_contraction_r}] and \emph{reduction} [Eq.~\eqref{eq_reduction}] rules, named after their resemblance to graph-merging operations~\cite{random-graph,graph-theor-appl}. Taken together, we arrive at $\alpha$-percolation, 
for which the new connection criterion is given by Eq.~\eqref{eq_criterion_mega}. 
The $\alpha$-percolation exhibits the following key features:}

{\subsection{Optimal order of conducting the graph rules}}
A key question in designing quantum communication strategies for large networks is determining the optimal order for applying the remote distillation and quantum relay protocols. We show that the following strategies are optimal within our framework of utilizing distributed memories (with more details given in the Appendix):

(1)While the quantum relay protocol can be repeatedly applied, we show that the optimal network connectivity is achieved by invoking Eq.~\eqref{eq_reduction} only once on a component $a$ immediately after $a$'s range is not large enough to reach any other component, $r_a<\min_b d_{ab}$~(Appendix~\ref{appendix_e}). This suggests that a component only acts as a relay \emph{after} its ability to merge with other components is exhausted. Until then, its focus should be on conducting remote distillation and establishing as many links as possible.

(2) We also show that the two graph rules can be applied to different nodes in \emph{any order}, leading to the same final state of optimal connectivity (Appendix~\ref{appendix_f}). This means that components can prioritize merging with their closest neighbors first, creating larger components. These larger components can then expand their reach to more distant nodes. Ultimately, this approach ensures that nodes are optimally connected in the final state. If certain nodes remain unconnected in the final state, no further improvements can be achieved by applying Eqs.~\eqref{eq_contraction_r}~and~\eqref{eq_reduction} in a different order.

\subsection{Positive feedback and hopping features}

The consecutive conduction of the graph rules immediately reveals two key features: firstly, a ``positive feedback,'' such that nodes within larger components achieve a greater range {[shown as bigger circles in Fig.~\ref{fig_demo}]}, therefore more likely to form even larger components;
secondly, ``hopping'' over components, such that two components of no adjacency can still be connected {[e.g.,~the blue bridge link in Fig.~\ref{fig_demo}]}. Note that analogous features have been separately studied in the percolation literature: 
while operating on distinct
mechanisms, positive feedback phenomena have been explored in explosive percolation~\cite{explos-percolation_dsn15,explos-percolation_dsggna19}, cascading failures~\cite{cascade-fail_bppsh10,netw-of-netw_gbsh12}, {and cumulative merging percolation~\cite{cumul-merge-percolation_ms16,cumul-merge-percolation_cps20,cumul-merge-percolation_ccpsc22}}, while the hopping feature has been investigated in tunneling continuum percolation~\cite{contin-percolation_fbcb14} and extended-range percolation~\cite{extend-percolation_cct23}. Yet, to our knowledge, the combined exploration of these two features has not been explored.

The positive feedback and hopping features become especially pronounced when $r_0$ approaches the percolation threshold,  $r_0^\text{th}$. At the threshold, the emerging giant component {(highlighted in blue)} tends to be more non-local and sparser as $\alpha$ increases [Fig.~\ref{fig_threshold}]. Intriguingly, only one parameter $\alpha$ enters the model and affects both $r_0^\text{th}$ and the distribution pattern of the giant component. In contrast, the number of memories per node, $m$, merely rescales $r_0$ [Eq.~\eqref{eq_r_0}] without altering the underlying percolation mechanism. Classical continuum percolation can be recovered as the $\alpha\to0$ limit.

\section{Real-World Fiber Network}

{First of all, it is worth noting that}
our theoretical framework is predicated on taking $\epsilon \to 0$ with $\epsilon d_0$ fixed {before} taking the thermodynamic limit $N\to \infty$. {In a more realistic scenario,} when considering finite $\epsilon$ and $d_0$, an additional length scale, $\beta = d_0 \ln 3$, becomes relevant, which effectively caps a node's enhanced range $r(s)$ such that it cannot exceed $\beta$. 
The physical reason behind this is that in Eq.~\eqref{eq_werner}, any distance $d_{ij}$ longer than $d_0 \ln 3$ 
suffers from 
{entanglement sudden death}~\cite{entanglement-sudden-death_ye04,entanglement-sudden-death_ye06,entanglement-sudden-death_talo10}, leading to absolute zero entanglement in the flying qubits.

We demonstrate potential practical applications by examining Sparkle's pan-European fiber network~\cite{sparkle}.
To reduce the length scale under $\beta$, we inserted full-fledged quantum repeaters~\cite{q-repeater_aeehjlt23} along longer cables as additional nodes. This modified the network topology by dividing cables into shorter, random segments that follow a Poisson distribution in length, averaged around $50$~km.
As Fig.~\ref{fig_real} shows, {with practically reasonable fidelity thresholds ($99\%$)} and quantum memory capacity, the enhanced model ($\alpha=\alpha^*$) forecasts over $90\%$ connectivity under the current laboratory limit of decoherence distance, $d_0\sim 300$~km~\cite{q-netw-fiber_imtat13}. This stands in stark contrast to using only local memories ($\alpha=0$), which would require a $d_0$ exceeding $600$~km. This comparison underscores the critical role of integrating distributed quantum memories in practical quantum communication networks~(Appendix~\ref{appendix_g}).

\section{Discussion}

Our work established the profound role of quantum memories on the connectivity of quantum communication networks through mapping to a percolation process with 
associated graph rules. A combination of the contraction and reduction rules also appears in variants of the strong-disorder renormalization group (SDRG)---an efficient spatial renormalization group technique to generate the ground state and low-energy excitations of heterogeneous quantum systems, especially in the vicinity of quantum phase transitions~\cite{q-phase-transit,sdrg_im05,sdrg_im18}.
As recently shown, positive feedback and hopping features also emerge in SDRG for systems
such as the Ising~\cite{sdrg-ising_mcw68,sdrg-ising_f92,sdrg-ising_ki10}, Heisenberg~\cite{sdrg-heisenberg_mdh79}, and Josephson junctions~\cite{sdrg-josephson_akpr04,sdrg-josephson_akpr10},
and often lead to exotic disordered universality classes and Griffiths phases. {It is an exciting future direction to explore the critical phenomena of $\alpha$-percolation through insights from SDRG. Since SDRG rules generally favour continuous transitions~\cite{sdrg-first-order_hhv12}, we expect that the phase transition of our model remains second-order for any $\alpha$.}

Moreover, given a component of size $s$, we do not have to use all $ms$ distributed memories but only a fraction of them, $ m s^{\eta}$, with $\eta\le 1$ acting as a tunable parameter. This adjustment effectively reduces the number of pairs to be distilled from $n$ to $n^{\eta}$, thus altering $\alpha$ to $\eta \alpha$. As a trade-off, the time complexity is also reduced to $f(n)=O(n^{\eta^2\log_2 n})$. This tunability opens up the possibility of adjusting $\alpha$ between $0$ and $\alpha^*$, striving for an optimal balance between efficiency and time complexity (rate). 
On the flip side, specific channels often have room for more efficient distillation protocols (e.g.,~$R$-state-based protocols~\cite{q-distill_rstedw18}), which could lead to a higher gain of fidelity and thus $\alpha>\alpha^*$. However, some specific protocols might also offer greater success probabilities $p$, which could diminish the apparent benefits of probabilistic parallelization using quantum memories. This interplay between fidelity and success probability in distillation presents an area for further investigation.

In practice, while state-of-the-art experimental memories already excel in individual metrics like conversion efficiencies or gate operation fidelities, no single device yet combines them all, making this a central engineering challenge for the entire field.
Yet, non-dynamic imperfections such as finite fidelity and efficiency can be readily absorbed into our model. We can map these noise sources onto an ``overhead'' $c$ for each link, $d_{ab}\to d_{ab}+c$, encapsulating the aggregate noise from these additional imperfections [Eq.~\eqref{eq_werner}]. In this view, a less-than-perfect memory simply increases the effective path length an entangled pair must traverse, naturally integrating into $\alpha$-percolation.

The finite coherence time of memories and the need for synchronization to manage timing jitter and latency also introduce a distinct temporal dimension that our model does not fully capture. We estimated the required coherence time for our non-dynamic model to remain approximately valid, finding a minimum requirement of $0.04$ seconds (Appendix~\ref{appendix_g}), which is in principle achievable with current NV nuclear spin technologies~\cite{q-mem_lwsdl23}. However, this analysis oversimplifies the interplay of other factors such as the reading/writing times. Integrating more practical factors requires a fully dynamic extension of our model in the future, which could offer more realistic performance predictions. Our goal here has been to present a so far overlooked fundamental mechanism that can help to achieve better connectivity in quantum communication networks even in more realistic settings.

\section*{Acknowledgment}

This work was supported by the National Science Foundation under Grant No.~PHY-2310706 of the QIS program in the Division of Physics. X.M. acknowledges the support by the Co-design Center for Quantum Advantage (C2QA).

\bibliographystyle{quantum}
\bibliography{refs}

\newpage
\onecolumngrid

\appendix

{\section{Entanglement Distillation Protocol (Point-To-Point)}}
\label{appendix_a}


{The primary use of quantum memories in quantum communication is for entanglement distillation protocols~\cite{q-distill_hh01}. The goal of a distillation protocol is to ``distill'' from many entangled pairs a single pair exhibiting higher entanglement. Here, we start by focusing on the BBPSSW protocol~\cite{entangle-purif_bbpssw96}, one of the most widely adopted distillation protocols.
We examine the protocol in the basic, 
\emph{point-to-point} scenario between two parties, Alice ($a$) and Bob ($b$), who share $2$ copies of an entangled pair of qubits described by the isotropic state:
\begin{eqnarray}
\label{eq_werner_si}
   \rho&=& p\left|\Psi^{-}\right\rangle\left\langle\Psi^{-}\right|+\left(1-p\right) \mathbf{I}/4,
\end{eqnarray}
where $\left|\Psi^{-}\right\rangle= \left(\left|01\right\rangle-\left|10\right\rangle\right)/\sqrt{2}$, and
\begin{equation}
\label{eq_p}
    p=e^{-d_{ab}/d_0}
\end{equation}
is a function of the minimum distance $d_{ab}$ between $a$ and $b$ relative to a characteristic length $d_0$. The fidelity of such a pair is calculated as $F=\left\langle\Psi^{-}\right|\rho \left|\Psi^{-}\right\rangle={\left(3p+1\right)}/{4}$~\cite{werner_w89}.
We also examined the DEJMPS protocol~\cite{entangle_purif_dejmps96}, which fully reduces to BBPSSW if using the isotropic state [Eq.~\eqref{eq_werner_si}] as the initial states.
}

{Following the distillation protocol,
$a$ and $b$ each apply a bilateral CNOT operation to the two qubits they own~\cite{entangle-purif_bbpssw96}.
In a CNOT operation, one qubit acts as a control, and the other as a target. If the control qubit is in the state $\left|1\right\rangle$, the target qubit is flipped; if the control qubit is in the state $\left| 0\right\rangle$, the target qubit remains unchanged.
After the CNOT operation, they measure the target qubits in the basis $\left| 0\right\rangle$ and $\left| 1\right\rangle$. If their measurement outcomes agree (both are $\left|00\right\rangle$ or $\left| 11\right\rangle$), the distillation succeeds and $a$ and $b$ keep the control qubits; otherwise, they discard the qubits. The success probability is given by
\begin{equation}
\label{eq_success}
    p=F^2+2F\left(1-F\right)/3+5 \left[\left(1-F\right)/3\right]^2.
\end{equation}
Once successful, $a$ and $b$ obtain a pair of control qubits that have higher fidelity:
\begin{equation}
    F'=\frac{F^2+\frac{1}{9}\left(1-F\right)^2}{F^2+\frac{2}{3}F\left(1-F\right)+
    \frac{5}{9}\left(1-F\right)^2}.
\end{equation}
In the limit $F\to 1$, one has $1-F'\simeq\left(2/3\right)\left(1-F\right)$.} 

{If $a$ and $b$ have $O(n)$ sufficient memories, then they can extend the BBPSSW protocol to $n$ bipartite qubit pairs
and implement the protocol
in a nested manner---initially with $n$ pairs, then with $n/2$ after the first iteration, and so on---allowing for a progressive enhancement of fidelity. The final fidelity will be given by:
\begin{equation}
    1-F_\text{final}\simeq\left(2/3\right)^{\log_2 n}\left(1-F\right),
\end{equation}
which can be reached in polynomial time. Note that although the same fidelity could be achieved without the use of memories (by implementing the BBPSSW protocol sequentially, not in a nested way), the time complexity is exponential, which is therefore not considered as achievable in large-scale quantum communication networks.}

{If the final pair's entanglement exceeds the fidelity threshold we set, $F_\text{final}\ge 1-\epsilon$, we treat this pair as connected by an \emph{(entanglement) link} in our continuum percolation model. This link can be further used for teleportation purposes.}

\newpage

\section{Remote Distillation}
\label{appendix_b}

\begin{figure}[h!]
	\centering
	\begin{minipage}[b]{121pt}	
		{\includegraphics[width=121pt]{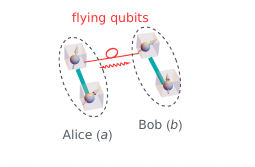}\vspace{-4mm}\subcaption{\label{fig_3d_a_si_4}}}
    \end{minipage}
    \begin{minipage}[b]{121pt}	
		{\includegraphics[width=121pt]{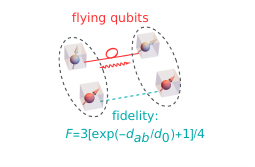}\vspace{-4mm}\subcaption{\label{fig_3d_a_si_5}}}
    \end{minipage}
    \begin{minipage}[b]{121pt}	
		{\includegraphics[width=121pt]{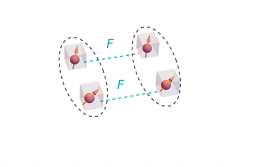}\vspace{-4mm}\subcaption{\label{fig_3d_a_si_6}}}
    \end{minipage}
    
    \begin{minipage}[b]{121pt}	
		{\includegraphics[width=121pt]{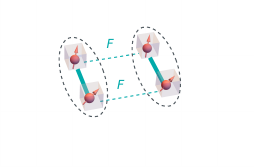}\vspace{-4mm}\subcaption{\label{fig_3d_a_si_7}}}
    \end{minipage}
    \begin{minipage}[b]{121pt}	
		{\includegraphics[width=121pt]{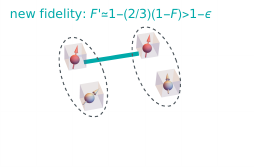}\vspace{-4mm}\subcaption{\label{fig_3d_a_si_8}}}
    \end{minipage}
    \begin{minipage}[b]{121pt}	
		{\includegraphics[width=121pt]{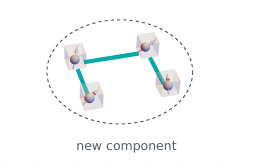}\vspace{-4mm}\subcaption{\label{fig_3d_a_si_9}}}
    \end{minipage}
    
    \caption{\label{fig_3d_a_si}Remote distillation protocol.
    \subref{fig_3d_a_si_4}~There are abundant flying qubits between the two nearest nodes of two components, Alice ($a$) and Bob ($b$). However, these flying qubits cannot be efficiently distilled without using memories. 
    \subref{fig_3d_a_si_5}~At the expense of existing entanglement links, the flying qubits are teleported to neighboring nodes and stored in the distributed memories.
    \subref{fig_3d_a_si_6}~{Several entangled pairs, represented by stored stationary qubits,} are formed in parallel between the two components.
    \subref{fig_3d_a_si_7}~The expended entanglement links are regenerated.
    \subref{fig_3d_a_si_8}~A new entangled pair of higher fidelity is distilled using remote gate teleportation (CNOT gates)---again, at the expense of existing entanglement links. The new pair can be treated as an entanglement link if the new fidelity is larger than the threshold, $F'>1-\epsilon$.
    \subref{fig_3d_a_si_9}~Existing entanglement links are regenerated. Now the two components merge into a larger component.
 \hfill\hfill}
\end{figure}

Traditional point-to-point distillation protocols assume the capability of Alice ($a$) and Bob ($b$) to apply CNOT operations across their respective qubits~\cite{q-distill_hh01}. 
However, in the following \emph{remote distillation} protocol, we consider Alice and Bob as representing connected components of spatially distributed nodes. Hence, Alice (Bob) is unable to directly perform a CNOT gate on qubits that are housed in different nodes within her (his) components, due to the principle that operations across separate nodes must adhere to the restrictions of local operations and classical communication (LOCC). The only way for Alice (Bob) to manipulate the qubits distributed across her (his) component and to perform a CNOT gate operation is through the technique of remote gate teleportation~\cite{q-remote-control_hvcp01}. The remote gate teleportation uses the connections---existing entanglement links---between different nodes within the component to teleport the CNOT gate operation from one node to another. This underpins the groundwork for the implementation of our remote distillation protocol.

The process is introduced as follows (Fig.~\ref{fig_3d_a_si}):

Firstly, consider two connected components, $a$ and $b$, each consisting of $s_a$ and $s_b$ nodes that are connected by existing links. 
Each link represents an existing, almost-perfectly entangled pair of qubits between two nodes in the component.
Our initial task involves identifying the two nearest individual nodes between the two components $a$ and $b$. The reason is that the channel between the two nearest nodes gives the highest fidelity of entanglement (for now) that can be established between $a$ and $b$. 

The next step is to distill entanglement from this identified channel. 
We assume that there are abundant flying qubits in the channel. This guarantees that there are always enough entangled flying qubits to be stored into all available memories.
The remote distillation protocol aims to employ not just the quantum memories housed in these two nearest nodes (assuming each node contains $m$ quantum memories), but also those within the entire component connected to these two nodes, effectively leveraging $ms_a$ and $ms_b$ quantum memories in total. To accomplish this, we first employ the existing entanglement links to teleport qubits from the two nearest nodes to other nodes within their respective components. The maximum number of entangled pairs can be stored in memories between the two components in parallel is determined by $n=\min\{ms_a,ms_b\}$. This teleportation step costs the entanglement links within the components, and thus one needs to wait for the regeneration of the links. Consequently, this regeneration introduces additional time complexity to the remote distillation protocol compared to traditional distillation protocols using local memories. The implications of this time complexity will be further discussed through a one-dimensional example in Section~VI.

Once the parallel storage of entangled pairs between components $a$ and $b$ is achieved, the subsequent phase involves teleporting and executing CNOT operations across all nodes in the two components that have entangled pairs stored. This action, again, costs the entanglement links within the network, thereby adding an additional layer of time complexity to the process.

Once our remote distillation protocol creates a new entanglement link, the link will merge previously separate components $a$ and $b$ into a larger component.
This process allows for the recursive application of the remote distillation protocol, thereby fostering a positive feedback (as discussed in the main text). Within the framework of continuum percolation theory, as we approach the limit of $\epsilon \to 0$, the positive feedback amounts to the increase of the reaching ``range'' of this larger component, which can be succinctly expressed as:
\begin{equation}
\label{eq_contraction_r_si}
    (r')^{1/\alpha}=r_a^{1/\alpha}+r_b^{1/\alpha},
\end{equation}
which is the \emph{contraction rule} highlighted in the main text.

\newpage
\section{Time Complexity of Remote Distillation in One Dimension}
\label{appendix_c}

As a proof-of-concept, here we consider remote distillation in one dimension, illustrating how strategic utilization of distributed memories can enhance the connectivity of a one-dimensional network---within \emph{subexponential time complexity}:

Denote $f(n)$ as the worst-case time complexity for establishing a perfectly entangled pair ($\epsilon\to 0$) within a component of size $\sim n$. Here, we assume $m=1$ w.l.o.g. The worst-case scenario corresponds to establishing a link near the \emph{mid point} of the component. Therefore, consider two components, each sized $\sim n/2$, and establishing a link through their nearest nodes.  Remote distillation requires utilizing memories not only within the two nearest nodes but also distributed across the respective components, mandating the employment of our introduced remote distillation scheme. 

Initially, the flying qubits are teleported from the nearest nodes to other nodes within the two components and stored in the distributed memory qubits. The total entanglement links spent for this task amounts to
\begin{equation}
    2\left[\left(n/2-1\right)+\left(n/2-2\right)+\cdots+1\right]\simeq \frac{1}{4}n^2,
\end{equation}
where $\left(n/2-1\right)$ is the number of links needed to teleport one flying qubit from the nearest node to the farthest node within one $n/2$-sized component; $\left(n/2-2\right)$ is to the second farthest node; and so on.
Consequently, the time complexity of rebuilding these links is approximately $O(n^2)f(n/2)$, where $f(n/2)$ denotes the worst-case time complexity of rebuilding a link within a component of size $n/2$.

Subsequently, teleportation of quantum gates~\cite{q-gate-teleport_jtsl07} (CNOT gates, specifically) are required to act on these memory qubits. The total number of links utilized for this purpose is:
\begin{equation}
    2\left[1\left(n/4\right)+2\left(n/8\right)+4\left(n/16\right)+\cdots+n/4\left(1\right)\right]\simeq \frac{1}{2} n \log_2 n.
\end{equation}
Here, $1(n/4)$ denotes the total count of entanglement links required during the first distillation round within one component, involving the teleportation of $n/4$ CNOT gates between every other pairs of nodes separated by a single link; for the second round, $2(n/8)$ specifies the requirement, with $n/8$ CNOT gates needing teleportation between nodes that are two links apart; and so on.
Consequently, the time complexity of rebuilding these links approximates to $O(n \log_2 n)f(n/2)$.

Finally, a careful manipulation of the remote distillation process directly produces the distilled high-fidelity pair between the two nearest nodes. Combining these operations yields a total time complexity $f(n)$ given by:
\begin{equation}
\label{eq_fn}
    f(n)\simeq \left(\frac{n}{2}\right)^{-\log_2 p}+ \left(\frac{1}{4}n^2+ \frac{1}{2} n\log_2 n\right) f(n/2),
\end{equation}   
where the first term represents the time complexity of parallel distillation using memories, while the second term accounts for the compensation of expended links for all remote teleportation purposes. Taking the logarithm of the equation and assuming $f(n) \gg n^{-\log_2 p}$, we derive:
\begin{equation}
\frac{\ln f(\ln n)-\ln f (\ln n -\ln 2)}{\ln 2}\simeq \frac{2}{\ln 2} \ln n,
\end{equation}   
which yields $\ln f \simeq \left(\ln 2\right)^{-1} \left(\ln n\right)^2$, or $f(n)=O(n^{\log_2 n})$, indicating subexponential time complexity.

Transitioning to higher dimensions, we anticipate even lower time complexity. This is since the depth of a component of size $n$ will be less than $n$ in higher dimensions. Consequently, the coefficient of the second term in Eq.~\eqref{eq_fn} could be smaller than $O(n^2)$. However, achieving this necessitates careful path routing of teleportation within the component.

In higher dimensions, $f(n)$ denotes the time complexity of establishing a connected component of size $\sim n$, which is usually {sparse}. To construct a {dense} component, at most $\sim n^2$ links are needed. Thus, the worst-case time complexity is pushed to $n^2 f(n)$, which remains subexponential in $n$. Typically, the time complexity can take different subexponential forms based on the required connectivity density of the component.

Finally, there is also a practical caveat: for effective utilization of entanglement links within a connected component at any given time, it is advantageous to store these links into memory qubits as well. Nonetheless, this only demands an additional constant number of memories. Therefore, these additional memories have no bearing on the asymptotic results discussed above.

\newpage
\section{Quantum Relay}
\label{appendix_d}

\begin{figure}[h!]
	\centering
	\begin{minipage}[b]{121pt}	
		{\includegraphics[width=121pt]{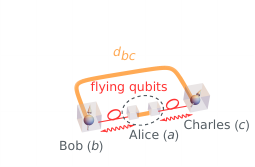}\vspace{-4mm}\subcaption{\label{fig_3d_b_si_4}}}
    \end{minipage}
    \begin{minipage}[b]{121pt}	
		{\includegraphics[width=121pt]{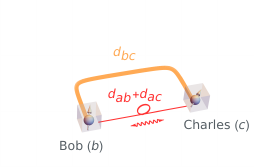}\vspace{-4mm}\subcaption{\label{fig_3d_b_si_5}}}
    \end{minipage}
    \begin{minipage}[b]{121pt}	
		{\includegraphics[width=121pt]{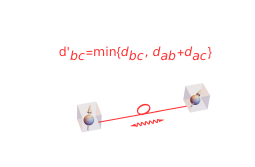}\vspace{-4mm}\subcaption{\label{fig_3d_b_si_6}}}
    \end{minipage}
    
    \caption{\label{fig_3d_b_si}Quantum relay protocol.
    \subref{fig_3d_b_si_4}~There are abundant flying qubits between Alice ($a$) and Bob ($b$), 
    as well as Alice ($a$) and Charles ($c$).
    \subref{fig_3d_b_si_5}~Entanglement swapping protocol on $a$ creates a direct channel between $b$ and $c$, with effective distance $d_{ab}+d_{ac}$.
    \subref{fig_3d_b_si_6}~The shortest channel produces the highest-fidelity flying qubits. Thus, only the shortest channel is kept.
 \hfill\hfill}
\end{figure}

A quantum relay protocol~\cite{q-relay_jpf02,q-relay_rmtzcg04} exclusively relies on Bell-basis entanglement swapping~\cite{entangle-swap_zzhe93,entangle-swap_sdsbbz05} applied to mixed states. The process is introduced as follows (Fig.~\ref{fig_3d_b_si}):

Consider three nodes, Alice ($a$), Bob ($b$), and Charles ($c$) in the network. Alice and Bob share a pair of flying qubits in the form of the isotropic state [Eq.~\eqref{eq_werner_si}], with fidelity $F_{ab}$, or $p_{ab}=\left(4F_{ab}-1\right)/3$. Similarly, Alice and Charles share a pair of flying qubits in the isotropic state, with fidelity $F_{ac}$, or $p_{ac}=\left(4F_{ac}-1\right)/3$. A successful Bell-basis  entanglement swapping requires a Bell-state measurement on the two flying qubits that Alice holds, followed by classical communication to Bob and Charles, which then leads to local transformations on Bob's and Charles's own qubits~\cite{entangle-swap_zzhe93}.
Given the decomposition of the isotropic state [Eq.~\eqref{eq_werner_si}], it can be shown that only with probability 
\begin{equation}
    p'_{bc}=p_{ab}p_{ac},
\end{equation}
or
\begin{equation}
\label{eq_swapping_f}
    \frac{4F'_{bc}-1}{3}=\frac{4F_{ab}-1}{3} \frac{4F_{ac}-1}{3},
\end{equation}
can a maximally entangled state be established directly between Bob and Charles. Otherwise, a completely mixed state is obtained. Considering the exponential dependence of $p$ on the distance [Eq.~\eqref{eq_p}], this equates to having an ``effective'' distance, $d_{ab}+d_{ac}$, between Bob and Charles.

Note that flying qubits are assumed to be abundant (as long as they can be constantly heralded~\cite{q-netw_hkmsvtmh18}). Therefore, we ignore the rate efficiency and focus on the quality (fidelity) of the flying qubits. Assume that there already exists a channel connecting $b$ and $c$, with distance $d_{bc}$.
Given two channels of (effective) distances $d_{bc}$ and $d_{ab}+d_{ac}$, the shorter one consistently yields higher-fidelity flying qubits. Consequently, only the {shorter} link needs to be retained. This leads to the \emph{reduction rule}:
\begin{equation}
\label{eq_reduction_si}
d'_{bc}=\min\{d_{bc}, d_{ab}+d_{ac}\}.
\end{equation}

At first glance, Eq.~\eqref{eq_reduction_si} might not appear to enhance fidelity. Suppose the nodes $a$, $b$, and $c$ are placed in a Euclidean space with all three quantum channels of Euclidean distances, $d_{ab}$, $d_{bc}$, and $d_{ac}$, established. The triangle inequality dictates $d_{bc}<d_{ab}+d_{ac}$, indicating that $d_{ab}+d_{ac}$ is always longer than the direct channel between $b$ and $c$. However, the enhancement is hidden in two scenarios. Firstly, if, for some reason, a channel between $b$ and $c$ cannot be established, then $d_{bc}\to \infty$. In this case, Alice, acting as the relay, will significantly reduce the effective distance between Bob and Charles to a finite value $d_{ab}+d_{ac}$. Another possibility arises when Alice is not a single node but represents a connected component, where the links are perfectly entangled (in the limit $\epsilon\to 0$). Consequently, the real distances of these perfect links become inconsequential, as the flying qubits can be perfectly teleported from one end to another. Thus, the effective distance $d_{ab}+d_{ac}$, disregarding the path routing length within component $a$, could actually be smaller than the real distance $d_{bc}$.

It is worth noting that the quantum relay operation operates independently of quantum memories and does not impact the overall time complexity. However, one might wonder about the potential utility of quantum memories within the component that serves as the relay. Here, we demonstrate that, to the first-order approximation ($\epsilon \to 0$), leveraging quantum memories within the isolated component does not enhance the quantum relay feature.

Consider the scenario where Alice possesses $m$ memories that can be utilized. This allows us to consider the following procedure: we initiate distillation between Alice and Bob, as well as between Alice and Charles, and then swap the two distilled pairs to establish a direct pair between Bob and Charles, the fidelity $F'_{bc}$ of this final pair (in the limit as $F\to 1$) is given by [Eq.~\eqref{eq_swapping_f}]:
\begin{equation}
    1-F'_{bc}\simeq \left(1-F_{ab}\right)+\left(1-F_{ac}\right),
\end{equation}
Here, $F_{ab}$ ($F_{ac}$) represents the fidelity of the two distilled pairs between $a$ and $b$ ($a$ and $c$):
\begin{eqnarray}
    1-F_{ab}&\simeq& \frac{3}{4} m^{-\alpha} \frac{d_{ab}}{d_0},\nonumber\\
    1-F_{ac}&\simeq& \frac{3}{4} m^{-\alpha} \frac{d_{ac}}{d_0}.
\end{eqnarray}
However, this new fidelity $F'_{bc}$ is identical to directly swapping $m$ pairs between Bob and Charles (through Alice) and then distilling them, resulting in a fidelity $F'$ given by
\begin{equation}
    1-F'\simeq \frac{3}{4} m^{-\alpha} \frac{d_{ab}+d_{ac}}{d_0}.
\end{equation}
A straightforward comparison yields $F'_{bc}=F'$. In other words, in the $\epsilon\to 0$ limit, distilling $m$ pairs between $a$ and $b$ ($a$ and $c$) and then swapping is \emph{equivalent} to swapping $m$ pairs from $b$ to $c$ (through $a$) and directly distilling them between $b$ and $c$. However, the latter approach offers the advantage of not requiring quantum memories within Alice.

\newpage
\section{Optimal Way to Apply Reduction Rule}
\label{appendix_e}

We claim that applying the reduction rule only on \emph{isolated} connected components can achieve the optimal connectivity. Here, an isolated component is defined as when the range of the component fails to reach any of the component's neighbors, i.e.,~$r_a<\min_b d_{ab}$. Additionally, we claim that it is only necessary to apply the reduction rule a single time to a component, immediately after it becomes isolated.

Consider an isolated component $b$ that has range $r_b$. It is clear that $r_b$ cannot grow larger. This is because by the definition of isolated components, for any neighbor $c$ of the component $b$, we must have $d_{bc}>r_b$.
Therefore, the $\alpha$-percolation criterion $d_{bc}<\min\{r_b, r_c\}$ will remain unsatisfied for $r_b$, inhibiting further growth of the size of $b$. The only exception is if a later reduction rule shortens the distance from $b$ to $c$, denoted as $d'_{bc}$, after acting on a common neighbor $a$ of both $b$ and $c$. 
Suppose that now $b$ can reach $c$. As a result, $r_b>d'_{bc}$. However, according to the reduction rule [Eq.~\eqref{eq_reduction_si}], this subsequently implies either $r_b>d_{bc}$ or $r_b>d_{ab}+d_{ac}> d_{ab}$, implying $b$ could reach either $a$ or $c$ \emph{before} reduction on $a$. This contradicts the initial assumption of $b$ being isolated.

Hence, we conclude that \emph{once a connected component becomes isolated, it remains isolated and cannot forge links with any other components}. 
Consequently, the reduction rule needs to be applied only once to any isolated component. This is because once a component is isolated, it will not merge with others or gain new neighbors, thus no new shortcuts will be generated.

After component $a$ is identified as isolated, it should be ``removed'' from the percolation process of the rest of the network after applying reduction to it. This removal means that when other nodes are tested whether they are isolated, they no longer count $a$ as a neighbor. As a result, other nodes may become isolated now after the removal of $a$. Ultimately, this process continues until all nodes are isolated, reaching the final state. Of course, this removal is only conceptual. It does not imply that nodes within isolated components are physically erased. The nodes are still linked with other nodes in the same isolated component.

Now we demonstrate that \emph{applying reduction to isolated components is more beneficial than applying it to non-isolated ones}. Note that a component always produces more efficient shortcuts when getting merged, so in general it is beneficial to wait until the component cannot merge any more, becoming isolated. The critical question, however, is whether reducing a component $a$ before it is isolated can offer a \emph{timely} advantage, potentially facilitating the merging of other components before those components become isolated. This scenario is impossible. If two components, $b$ and $c$, could benefit from a shortcut $d'_{bc}$ created by $a$, then it must be that $r_b>d'_{bc}$ and $r_c>d'_{bc}$. This implies that either both $r_b>d_{bc}$ and $r_c>d_{bc}$ are true, indicating $b$ and $c$ do not require $a$'s shortcut, or both $r_b>d_{ab}+d_{ac}>d_{ab}$ and $r_c>d_{ab}+d_{ac}>d_{ac}$ are true, suggesting that $b$ and $c$ cannot be considered isolated before $a$ is removed. Therefore, if $b$ and $c$ need a shortcut from $a$, they can afford to wait until $a$ is isolated. Of course, once $a$ becomes isolated, the reduction rule should be applied immediately on $a$, before $a$ is removed. Given that all nodes will eventually become isolated, this principle applies universally.

\newpage
\section{Sequence of Conducting the Graph-Merging Rules}
\label{appendix_f}

Now we investigate whether the sequence of implementing the two graph-merging rules holds significance, arguing that, in fact, the order does \emph{not} affect the outcome:

(1) The \emph{contraction rule}, as expressed by Eq.~\eqref{eq_contraction_r_si}, is commutative. Consequently, the enhanced range $r(s)$ of a node within a component solely depends on the component size $s$, irrespective of the specific order of contracting different nodes within the component. The whole process is therefore equivalent to a component-finding algorithm, which can be simply done using, e.g.,~a depth-first search. Also, note that the distance between two components $a$ and $b$ is defined as the distance between their nearest nodes, i.e.,~$d_{ab}=\min_{i\in a, j\in b}{d_{ij}}$. This again does not depend on the specific order of applying the contraction rule to the two components either.
Consequently, these two properties together establish the effective-node picture, where a connected component can be equivalently regarded as an effective node with range $r(s)$.

(2) Similarly, the minimum operation in the \emph{reduction rule} [Eq.~\eqref{eq_reduction_si}] is also commutative. Thus, the sequence in which isolated effective nodes are reduced also does not influence the outcome. In fact, given an isolated effective node $a$, we do not even actually need to remove $a$ and add  a ``shortcut'' $d'_{bc}$ between every pair of its neighbors $b$ and $c$. Instead, we can simply run a shortest-path algorithm, such as Dijkstra's algorithm, to identify the shortest path length between $b$ and $c$, whether through the relay $a$ or not. When there are multiple isolated effective nodes, this shortest-path approach can significantly reduce the computational complexity by avoiding the potential overload of shortcuts established throughout the network.

(3) The only remaining point that requires investigation is when both the contraction and reduction rules are involved. 
Given the requirement that reduction only happens for isolated components, we have posited that this approach yields optimal connectivity. Then, the question becomes whether altering the sequence of these rules might prevent certain nodes, say $i$ and $j$, which would otherwise merge optimally, from being able to do so. This scenario, however, is impossible. If $i$ and $j$ are merged in the final state, they cannot become isolated mid-process, as a continuous path for their eventual convergence always exists, which means $i$ will always have a neighbor $k$ such that $r_i>d_{ik}$, hence $i$ cannot be isolated. The only exception is that $d_{ik}$ was originally larger than $r_i$ but was later shortened by a reduction rule on $i$ and $k$'s common neighbor, say $l$.  Yet, according to Eq.~\eqref{eq_reduction_si}, this situation still ensures that $r_i>d_{il}+d_{lj}>d_{il}$. Thus, under no circumstances can $i$ become isolated. The same argument also works for $j$. And finally, contraction on $i$ (or $j$) with other nodes also does not impact $i$ and $j$'s connectivity: due to the ``minimum rule'' definition of the distance between two components, $d_{ab}=\min_{i\in a, j\in b}{d_{ij}}$, node $i$'s shortest distance to $j$ will only decrease after merging with other nodes. Eventually, $i$ and $j$ must be merged in the final state.

Drawing from these observations, we argue that the sequence in which the contraction and reduction rules are applied is inconsequential. The final state of configuration---how nodes are distributed across components---remains unchanged. Our algorithm initially applies contraction rules to the maximum number of nodes feasible, then proceeds to reduce all isolated components to add shortcuts. This cycle is repeated until the final configuration is established.

Note that for many traditional SDRG rules, the sequence of applying the rules plays a crucial role~\cite{sdrg_k12}. In SDRG, different merging orders of sites and bonds could often yield distinct outcomes. Moreover, SDRG rules are typically more approximate away from the critical point. This is in contrast to our $\alpha$-percolation rules, which are more generic, applicable regardless of whether at criticality or not, and not influenced by the order in which they are performed.

\newpage
\section{Detailed Analysis on Real Quantum Networks}
\label{appendix_g}

\begin{figure}[h!]
	\centering
    \begin{minipage}[b]{110pt}	
		{\includegraphics[width=110pt]{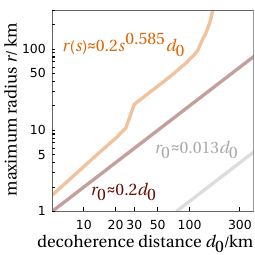}\subcaption{\label{fig_real_r}}}
    \end{minipage}
    \begin{minipage}[b]{220pt}	
		{\includegraphics[width=220pt]{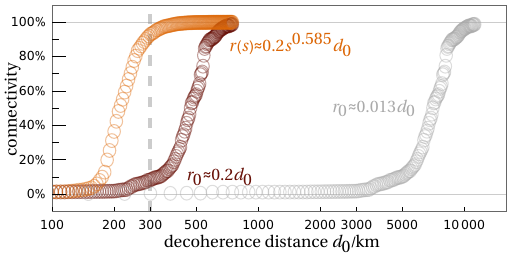}\subcaption{\label{fig_real_all}}}
    \end{minipage}
	
	\caption{{Different memory utilization strategies implemented on Sparkle's pan-European fiber network.
\subref{fig_real_r}~From right to left: absence of memory (gray), point-to-point memory utilization (brown), and distributed memory utilization (orange). Enabling distributed memories enhances the maximum range achievable by a node.
\subref{fig_real_all}~Correspondingly, 
in the absence of quantum memories (gray), the decoherence distance $d_0$ must surpass $\sim 8000$~km to ensure sufficient connectivity (above $90\%$) among nodes. Incorporating quantum memories already yields substantial improvements at the  point-to-point level (brown), reducing the required $d_0$ to just $\sim 600$~km. Our model further enhances this by leveraging distributed quantum memories at the component level (orange), cutting down the necessary $d_0$ to $\sim 300$~km---a distance now within current laboratory capabilities. The fidelity bound for each link is set at $1-\epsilon=99\%$. Each node is assigned $m \approx 102$ quantum memories such that $m^{\alpha^*}\approx m^{0.585}=15$.  The comparison between point-to-point versus distributed memory usage is also illustrated in Fig.~3(b) of the main text.}
    \hfill\hfill}
\end{figure}

The original Sparkle's pan-European fiber network~\cite{sparkle} has an average length scale of approximately $\sim 500$ kilometers, surpassing the current laboratory limit of $\beta = d_0 \ln 3 \approx 330$ kilometers. Due to \emph{entanglement sudden death}~\cite{entanglement-sudden-death_ye04}, it is impossible---via any combination of quantum communication protocols---to establish entanglement across such a typical length scale within the original fiber network. 
This situation is reminiscent of the role of finite-temperature for quantum phase transitions, where the inverse temperature $\beta$ introduces a finite length scale, capping the correlation length~\cite{q-phase-transit}.
Hence, the introduction of quantum repeaters~\cite{q-repeater_aeehjlt23} becomes necessary to reduce the length scale. To be specific, we altered the topology of the original fiber network by segmenting cables into shorter, randomly distributed segments following a Poisson distribution in length, with an average around $50$ kilometers. These newly introduced nodes, which concatenate these segments, effectively function as full-fledged quantum repeaters. Consequently, the resulting network comprises $692$ nodes (including the repeaters) and $733$ links. For any pair of nodes $i$ and $j$ not directly connected by a cable, we set $d_{ij}\to \infty$; otherwise, $d_{ij}$ is the cable length between $i$ and $j$.

In our simulations, we assumed that each node was equipped with approximately $m \approx 102$ quantum memories, yielding $m^{\alpha^*}\approx m^{0.585}=15$. The fidelity bound was set at $1-\epsilon=99\%$. We examined three scenarios: 

{\begin{enumerate}
\item No memory utilization, where each node's range $r_0\simeq 4\epsilon d_0 /3\approx 0.013 d_0$ remains fixed. 
Continuum percolation requires comparing $d_{ij}$  for every pair of nodes $i,j$, forming a link between $i$ and $j$ if the criterion $d_{ij}<r_0$ is met.
\item Point-to-point memory utilization, resulting in an increased range for each node to $ r_0 \simeq 4 m^{\alpha^* }\epsilon d_0 /3\approx 0.2 d_0$. The number of memories $m$ is included to account for their advantages in traditional point-to-point distillation protocols. The criterion of connection remains the same.
\item Distributed memory utilization, facilitated by the synergy between quantum communication protocols at the component level, leading to a size-dependent range $r(s)\simeq 4\epsilon \left(m s\right)^{\alpha^* }d_0/3 \approx 0.2 s^{0.585} d_0 $. The total number of remote memories $ms$ is included to capture their advantages for remote distillation, utilizing all memories within the component.
Therefore, once two components $a,b$ meet the criterion $d_{ab}<\min\{r_a, r_b\}$ and a link forms between them, the new component, of size $s_a+s_b$ will increase all its nodes' range [previously either $r(s_a)$ or $r(s_b)$] to a new, increased range $r(s_a+s_b)=0.2 \left(s_a+s_b\right)^{0.585} d_0$.
\end{enumerate}
}

As $d_0$ increases, Fig.~\ref{fig_real_r} illustrates the distinct behaviors for the three scenarios. Notably, as nodes are merged into larger components, the maximum $r(s)$ (i.e.,~of the largest component) experiences a nonlinear increase. This is since 
increasing $d_0$ leads to larger component sizes $s$, which subsequently results in a greater range $r(s)$. This enhanced range, in turn, contributes to better network connectivity, as shown in Fig.~\ref{fig_real_all}.

{
\emph{Time complexity analysis.---}As a figure of merit, we consider the worst-case success probability of the BBPSSW protocol, which is given by $p=F_\text{worst}^2+2F_\text{worst}\left(1-F_\text{worst}\right)/3+5 \left[\left(1-F_\text{worst}\right)/3\right]^2$ [Eq.~\eqref{eq_success}]. 
In the expression, the worst-case fidelity 
is given by $1-F_\text{worst}\simeq \left(3/4\right) \left(d_\text{worst}/d_0\right)\approx 0.25$, where $d_\text{worst}\sim 100$~km is the maximum distance for establishing entanglement to achieve $90\%$ connectivity in the fiber network. This gives rise to $p\approx 0.722$. 
}

Now, we put $p$ into the time complexity expression [Eq.~\eqref{eq_fn}]. While Eq.~\eqref{eq_fn} is for one dimension, we expect that the time complexity for higher-dimensional systems will not exceed Eq.~\eqref{eq_fn}.
When $m>1$, 
the time complexity of remote distillation of $n$ pairs from $n/m$ nodes is given by
\begin{equation}
\label{eq_fn_m}
    f(n)\simeq \left(\frac{n}{2}\right)^{-\log_2 p}+ m \left(\frac{1}{4}\frac{n^2}{m^2}+ \frac{1}{2} \frac{n}{m} \log_2 \frac{n}{m}\right) f(n/2).
\end{equation} 
By applying $m\approx 102$, we derive
\begin{eqnarray}
    f(102)&\approx& 7\nonumber\\
    f(204)&\approx& 1.3\times 10^3 \nonumber\\
    f(408)&\approx& 1.1\times 10^6 \nonumber\\
    &\vdots&
\end{eqnarray}
Although this subexponential increase is significant, within the fiber network, the maximum number of remote memory accesses needed is only given by
\begin{equation}
    n \simeq \left(\epsilon^{-1}\frac{3d_\text{worst}}{4 d_0}\right)^{1/\alpha}\approx 245,
\end{equation}
and thus by interpolation, $f(245)\approx 1.2 \times 10^{4.5}\approx 4 \times 10^{4}$. Therefore, assuming a photon detection rate of $10^6$~Hz for every channel, the required coherence time of quantum memories is given by $f(245)\times 10^{-6}\approx 0.04$~seconds, which is the minimum time needed to store all photons before they are used by the remote distillation protocol. This value is already achievable with current NV nuclear spin technologies~\cite{q-mem_lwsdl23}.

It is worth noting, however, that this analysis oversimplifies the interplay of many other factors, notably the reading/writing times of quantum memories, conversion efficiencies, gate operation fidelities, and, most crucially, photon transmission losses. Note that the transmission losses over long distances can be mitigated by scaling up the number of photon emission resources per unit time, though this may come at a significant financial cost.


\end{document}